\documentclass[prb,superscriptaddress,aps,twocolumn,10pt,showkeys]{revtex4-1}
\usepackage{graphics,graphicx,epsfig,color,latexsym,float,amsmath,ulem}
\usepackage[colorlinks,citecolor=blue,linkcolor=blue,urlcolor=blue]{hyperref}
\usepackage[version=4]{mhchem}
	
\newcommand{\avg}[1]{\left< #1 \right>} 
\newcommand{\ket}[1]{\left| #1 \right>} 
\newcommand{\bra}[1]{\left< #1 \right|} 

\begin{document}
\title{Atomic scale Skyrmions and large topological Hall effect \\in a breathing-kagome lattice}
\author{Nyayabanta Swain}
\email{nyayabanta@gmail.com}
\affiliation{Centre for Advanced 2D Materials, National University of Singapore, 6 Science Drive 2, Singapore 117546}
\affiliation{MajuLab, International Joint Research Unit IRL 3654,
CNRS, Universit{\' e} C{\^ o}te d’Azur, Sorbonne Universit {\' e},
National University of Singapore, Nanyang Technological University, Singapore}
\author{Munir Shahzad}
\email{munir.shahzad00@gmail.com}
\affiliation{Department of Physics and Atmospheric Science, Dalhousie University, Halifax, Nova Scotia, Canada, B3H 4J5}
\author{Pinaki Sengupta}
\email{psengupta@ntu.edu.sg}
\affiliation{School of Physical and Mathematical Sciences, Nanyang Technological University 637371, Singapore}

\date{\today}

\begin{abstract}
Motivated by recent experiments in \ce{Gd3Ru4Al12}, 
we demonstrate the emergence of atomic scale Skyrmions in interacting
spins on a breathing kagome lattice with competing nearest neighbor 
ferromagnetic and next nearest neighbor antiferromagnetic
exchange interactions. In the presence of an applied longitudinal magnetic
field, the ground state magnetic order evolves from a  helical phase at low
fields to a Skyrmion phase at intermediate fields before finally entering
a polarized phase at high fields. The size of each Skyrmion spans only two
unit cells of the lattice, in contrast to tens to hundreds of unit cells in 
most chiral magnets. Furthermore, the Skyrmions 
are driven not by chiral interactions but by the interplay between competing
exchange interactions and geometric frustration, just as 
in \ce{Gd3Ru4Al12} . When itinerant electrons
are coupled to the localized moments, they exhibit the
usual Skyrmion-driven topological Hall effect (THE) arising 
from the real space Berry curvature of the the Skyrmion texture. 
The small size of the Skyrmions in this system yield a strong local Berry 
curvature that results in an enhanced THE, which is investigated 
using a strong coupling approximation between the spins of the
itinerant electrons and the localized moments. Our results will
be crucial in understanding the experiments in \ce{Gd3Ru4Al12}
and other members of the same family of metallic frustrated magnets. 
\end{abstract}

\maketitle

\section{Introduction}
\label{sec:intro}
	

The observation of atomic scale Skyrmions and 
associated large topological Hall effect (THE) 
in centrosymmetric, Gd-based intermetallic compounds 
marks a significant development in the rapidly growing field of
Skyrmionics~\cite{Gd_expt1,Gd_expt2,Gd_expt3,Gd_expt4,Gd_expt5}. 
In magnetic systems, Skyrmions arise as spontaneously formed 
spin textures with non-trivial topology~\cite{Bogdanov_theory_analytical_2001,rosler_spontaneous_2006,nagaosa_topological_2013}.
These spin textures are topologically protected against any defects~\cite{nagaosa_topological_2013,hagemeister_stability_2015} 
making them attractive for practical applications such as 
magnetic data storage and processing~\cite{Wiesendanger_skyrmion_application_2016,Fert_skyrmion_application_2017}. 
At the same time, coupling of electric currents 
to these spin textures lead to large spin-transfer torques 
that arise at very low current densities~\cite{jonietz_spin-torque_2010,fert_spin-torque_2013} 
and unusual magneto-electric phenomena~\cite{wang_strong_2015,gobel_magnetoelectric_2019} 
such as topological Hall effect ~\cite{Neubauer_THE_MnSi_2009,kanazawa_MnGe_2011}.


Until recently, magnetic Skyrmions have been observed in 
$(i)$~non-centrosymmetric chiral magnets 
such as MnSi, FeGe, etc. where they arise from 
the interplay between Heisenberg exchange and 
intrinsic Dzyaloshinskii-Moriya interaction (DMI)~\cite{Neubauer_THE_MnSi_2009,yu_FeGe_2011,wilhelm_FeGe_2011,Nagaosa_theory_MC_2009,Nagaosa_theory_analytical_2010,theory_MC_Ambrose_2013,theory_continum_2016,theory_simulation_Krauth_2019,theory_simulation_Dagotto_2019},
and $(ii)$~heterostructures of magnetic thin films and heavy metals
where they are stabilised by interfacial DMI 
induced by strong spin-orbit coupling in the heavy metal layer~\cite{soumyanarayanan_skm_multi-layer_2017,raju_skm_multi-layer_2019,theory_MC_DFT_Dupe2014,theory_MC_DFT_Bottcher_2018}. 
Their typical size ranges from 10-100 nm for chiral magnets
and 100 nm - 10$\mu$m for magnetic thin films~\cite{Neubauer_THE_MnSi_2009,yu_FeGe_2011,wilhelm_FeGe_2011,soumyanarayanan_skm_multi-layer_2017,raju_skm_multi-layer_2019}. 
Atomic scale Skyrmions in Gd-based intermetallics 
are remarkable for their significantly smaller size (2.8 nm 
in \ce{Gd3Ru4Al12}) and, consequently, 
even more attractive for spintronic applications 
as they can be driven by a much smaller current 
and lower energy loss~\cite{Wiesendanger_skyrmion_application_2016,Fert_skyrmion_application_2017}. 
Skyrmions in these centro-symmetric materials are believed to be stabilized by longer range interactions~\cite{mechanism_skyrmion_Gd_mats,skyrmion_Gd_mats_RKKY_mechanism} -- a
marked departure from the DMI-driven Skyrmions.
These compounds have attracted 
widespread interest for both practical and fundamental reasons~\cite{Motome_theory_square_Gd_skyrmions}.
However, given their recent discovery, 
our knowledge about these materials remain largely incomplete,
including an understanding of the detailed mechanism of 
the formation of competing finite range interactions 
and magneto-transport properties.


In this work, we investigate a minimal microscopic model 
to understand the emergence of atomic scale Skyrmions 
from finite range competing exchange interactions 
in a centrosymmetric lattice, inspired by experiments on \ce{Gd3Ru4Al12}.
\ce{Gd3Ru4Al12} is a quasi-2D centrosymmetric 
metallic frustrated magnet where the localized 
Gd {\it 4f} and {\it 5d} orbitals form static magnetic moments 
while delocalized Ru-{\it 4d} orbitals contribute 
the itinerant electrons~\cite{mechanism_skyrmion_Gd_mats}. 
The magnetic moment carrying Gd$^{3+}$ ions are arranged 
in a breathing kagom{\' e} lattice geometry (fig.1) in each layer.
The coupling between the localized moments and 
itinerant electrons result in unique magnetic ordering and
magneto-transport properties, which is the focus of this work. 
It should be emphasized that an accurate explanation of 
every feature of the magnetic and transport measurements 
is beyond the scope of the current study. 
Instead we strive  to capture the principal features of 
the magnetic phase diagram and magneto-transport experiments.
We focus on various non-collinear magnetic phases stabilized on 
this lattice and their effect on the conduction electron motion 
via a coupled electron-spin model~\cite{Nagaosa_SkX_THE_2015,Mertig_SkX_THE_2017,gobel_family_2018,gobel_overcoming_2019,THE_SSL_no_skyrmion,THE_SSL_skyrmion}. 
Our results can be summarized as follows: 
$(i)$~Competing interactions give rise to a 
Skyrmions phase in this lattice. Unlike the Skyrmions 
in chiral magnets, the Skyrmions in our model 
are found to be of atomic sized scales.
$(ii)$~The topological Hall conductivity shows distinct features
revealing the emergence of topological electronic bands 
due to coupling to the localized moments. Crucially, 
our results demonstrate that competing exchange interactions 
are responsible for the appearance of Skyrmions 
in centrosymmetric lattices.

The rest of the paper is organized as follows. 
In section \ref{sec:model} we discuss the models used in this study.
In section \ref{sec:methods} we describe the method and the observables
we calculate. Section \ref{sec:results} contains the detailed results
of our work, followed by the summary in section \ref{sec:summary}.


\section{Model}
\label{sec:model}
	
\begin{figure}[b]
\centering
\includegraphics[width=6.0cm,height=5.0cm]{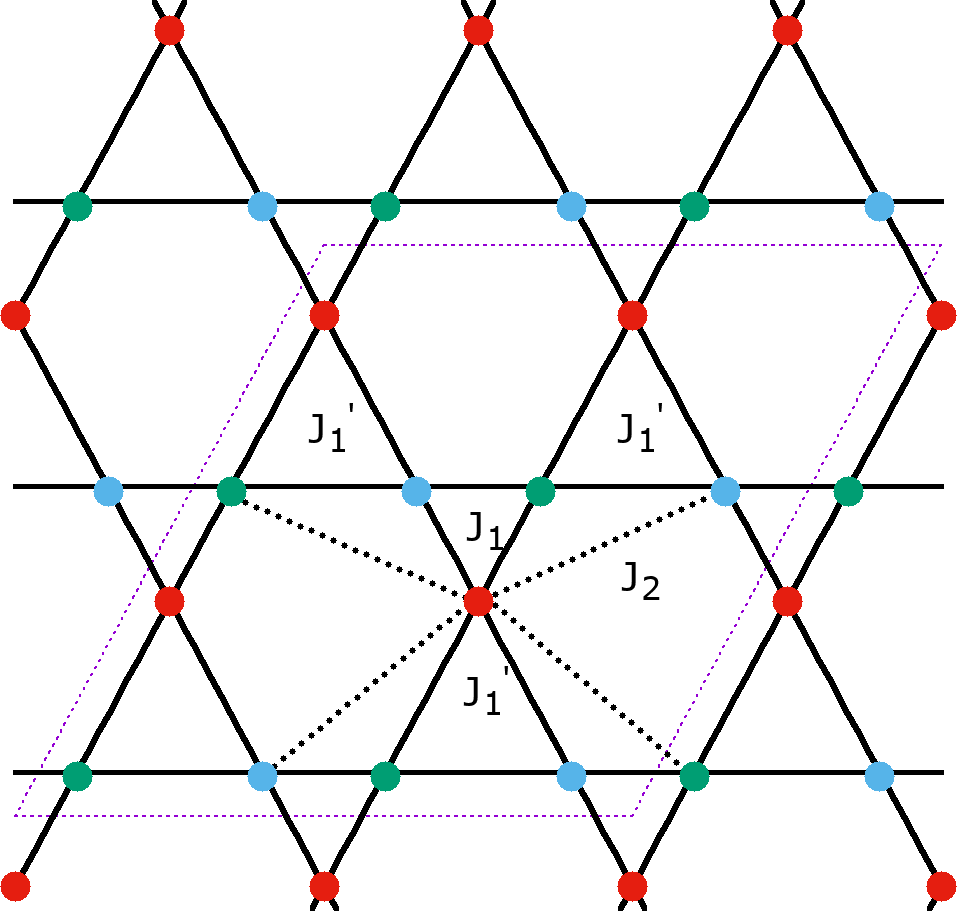}
\caption{\label{lattice}
Color online: Schematic diagram of a breathing kagome lattice 
with the nearest-neighbour (NN) Heisenberg exchange interactions
($J_1$ and $J'_1$) and the next-nearest-neighbour (NNN) 
Heisenberg exchange interaction ($J_2$) considered in our study.
The highlighted region involving 12 sites corresponds to the unit cell
for a Skyrmion lattice configuration (see Section \ref{sec:results} for details).
}
\end{figure}

We start by constructing minimal microscopic Hamiltonian to describe the 
localized moments. The dominant interaction between the spin degrees of freedom
is the Heisenberg exchange interaction. The differing bond lengths for the 
``up" and ``down" triangles in the breathing kagom{\' e} lattice geometry
is reflected in the differing strengths of nearest neighbor exchange interaction
on the corresponding bonds (see Fig. \ref{lattice}). Driven by the experimental observations
and first principle calculations in \ce{Gd3Ru4Al12}, we choose competing
nearest neighbor ferromagnetic and next nearest neighbor antiferromagnetic
Heisenberg interactions and an easy-plane single-ion anisotropy~\cite{mechanism_skyrmion_Gd_mats}. The absence 
inversion symmetry breaking forbids any Dzyaloshinskii-Moriya interaction. 
The complete magnetic Hamiltonian is given by,
	
\begin{align}
\label{equ:ham-class}
H_{cl} = & -J_1 \sum_{\langle i,j \rangle \in \Delta}  [S_i^x S_j^x + S_i^yS_j^y + \alpha S_i^zS_j^z ] \nonumber \\
	     &  -J_1' \sum_{\langle i,j \rangle \in \nabla}  [S_i^x S_j^x + S_i^yS_j^y + \alpha S_i^zS_j^z ] \nonumber \\
	     &  +J_2 \sum_{\langle\langle i,j \rangle \rangle} [S_i^x S_j^x + S_i^yS_j^y + \alpha S_i^zS_j^z ] \nonumber \\
	     & + A \sum_{i} (S_i^z)^2 - B \sum_{i} S_i^z 
\end{align}

\noindent where $J_1, J_1'$ are the ferromagnetic (FM) nearest-neighbour (NN) 
and $J_2$ is the anti-ferromagnetic (AFM) next-nearest-neighbour (NNN) 
Heisenberg exchange interactions  ($J_1'/J_1 = 0.75$). 
$\alpha = 1.5$ is the exchange anisotropy.
$A = 0.5J_1$ is the on-site easy-plane anisotropy, and
$B$ represents the Zeeman coupling of the spin 
with the external magnetic field applied along the $z-$axis.
Since Gd$^{3+}$ ions carry a large moment (${\bf S}=7/2$), the 
localized spins can be treated as classical vectors.
	
The itinerant electrons are modelled by a single orbital on each
lattice site with delocalization between nearest neighbors described by
hopping amplitudes that are proportional to the bond lengths. The electrons 
are coupled to the localized moments via a Kondo-like coupling term. 
Collecting all the terms, the transport properties of conduction electrons 
on the background of localized spin textures is described by the Hamiltonian

\begin{align}
\label{equ:ham-elec}
H &=  H_{el} +  H_{coup} \\
H_{el} &=  -t \sum_{\langle i,j \rangle \in \Delta} c_{i\sigma}^\dagger c_{j\sigma}
	     -t' \sum_{\langle i,j \rangle \in \nabla} c_{i\sigma}^\dagger c_{j\sigma}
	     -\mu \sum_i n_i  \nonumber \\
H_{coup} &= J_K \sum_i {\bf S}_i \cdot {\bf s}_i  \nonumber 
\end{align}
where $H_{el}$ is the kinetic energy, 
involving nearest neighbor hoppings $t$, $t'=0.75t$
on the breathing-kagom{\' e} geometry (see fig.\ref{lattice}).
$c_{i\sigma}^\dagger$ ($c_{i\sigma}$)
is a creation (annihilation) operator 
of an itinerant electron at site ${\bf r}_i$ with spin $\sigma$.
The chemical potential, $\mu$ controls the density of electrons.
The final term is the Kondo interaction, which describes 
the coupling between the local moment ${\bf S}_i$ 
to the electron spin operator ${\bf s}_i = 
\sum_{\alpha \beta} c^{\dagger}_{i \alpha} {\vec \sigma}_{\alpha \beta} c_{i \beta}$ 
at each site $i$.

\section{Method and observables}
\label{sec:methods}

\begin{figure*}[t]
\centering
\includegraphics[width=17.0cm,height=4.0cm]{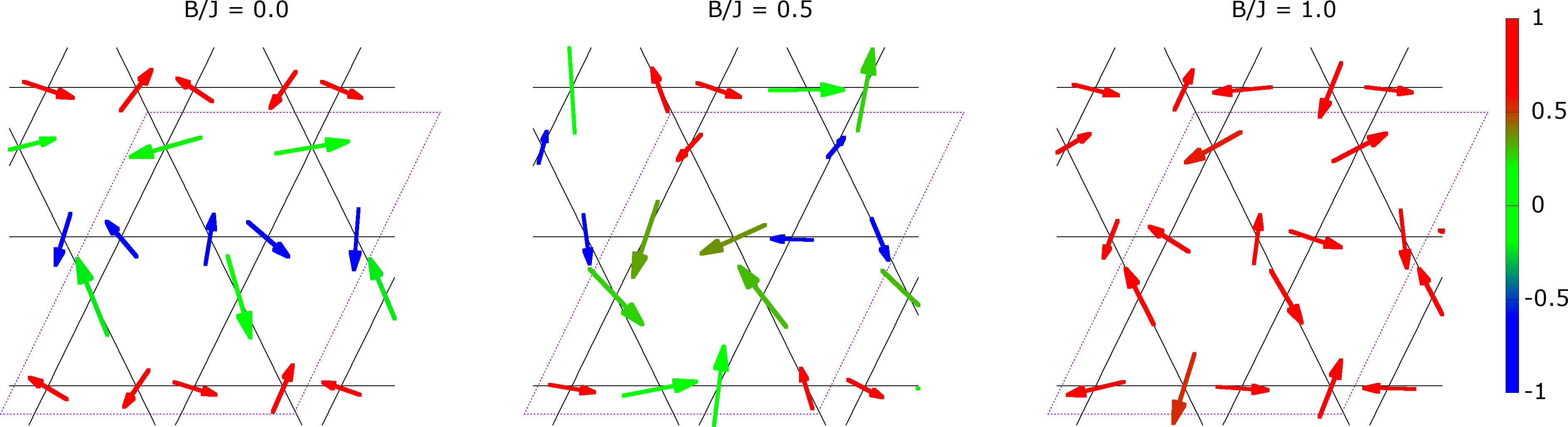}
\caption{\label{config1}
Color online: Representative spin configurations 
in the ground state obtained for different Zeeman coupling 
from our Monte-Carlo simulation.
The $xy$-components are represented by arrows in the $xy$ plane, 
whereas the $z$-component is represented by the color scale.
}
\end{figure*}

Given the large size of the moments, the dynamics of the magnetic
degrees of freedom are much slower than that of the itinerant
electrons and the two can be decoupled without affecting the low
energy physics. We make a further simplification by assuming that
the magnetic phases are determined exclusively by $H_{cl}$. 
While studying the electronic transport properties, 
the magnetic order is treated as static.
We study Hamiltonian (\ref{equ:ham-class}) with classical Monte 
Carlo~(MC) simulation on finite systems of size 
$3 \times L \times L$ ($L=$ 24 and 36) with periodic boundary conditions. 
Efficient thermalization is ensured by a simulated annealing procedure,
where the MC simulation is started from a random spin configuration 
corresponding to high temperature ($T_{high} \sim 1$), 
and then the temperature is reduced in steps of 
$\Delta T =0.01$ to reach the lowest temperature of $T_{low} = 0.01$, 
equilibriating the system at each temperature.
The equilibrium state at $T_{low}$ is used as the ground state 
for calculating physical observables.
At each temperature, we use $5 \times 10^5$ MC sweeps for equilibration,
and another $5 \times 10^5$ MC sweeps (in steps of 5000 sweeps) 
for calculating the observable in the ground state.
Near the phase boundary, we start with the equilibrated ground state, 
and heat the system using the Monte Carlo method to study its finite temperature properties. 
As a result the meta-stable phases have been avoided in this regime.

Using the above approach, we explore the magnetic phase
diagram of our model in the parameter space of magnetic field and temperature.
We compute several physical observables such as magnetization,
static spin structure factor and scalar spin chirality 
to characterize the different magnetic phases and identify 
the intervening phase transitions. The different magnetic
orderings are further confirmed by the
real space configurations of the localized spins. 

The long range magnetic ordering is identified by the static spin structure factor, defined as the Fourier transform of the 
equal-time spin-spin correlation.
\begin{align}
S({\bf Q})=\frac{1}{N}\sum_{i,j}\avg {{\bf S}_i \cdot {\bf S}_j} 
\exp [i{\bf Q} \cdot ({\bf r}_{i} - {\bf r}_{j})]
\end{align}
The spin textures are further characterized by the spin chirality, defined as
\begin{align}
\chi = \frac{1}{4\pi N} \langle \sum_{\langle ijk \rangle} 
[{\bf S}_i \cdot ({\bf S}_{j} \times {\bf S}_{k})] \rangle,
\end{align}
where the triple product is calculated for each traingular plaquette.
A non-zero $\chi$ signifies a non-coplanar spin texture and is crucial 
in distinguishing the Skyrmion phase from the helical and fully polarized phases.

The effects of coupling between localized moments and itinerant 
electrons on magneto-transport is investigated by calculating the 
transverse conductivity from Hamiltonian (\ref{equ:ham-elec}) using 
the Kubo formula~
\begin{align}
\sigma_{xy}=\frac{ie^2\hbar}{N}\sum_{m,n\neq m}(f_m-f_n) \frac{\bra{m}v_x\ket{n}\bra{n}v_y\ket{m}}{(\mathcal{E}_m-\mathcal{E}_n)^2+\eta^2}
\end{align}

\noindent where indices $m$ and $n$ represent the sum over all energy levels, 
$N$ is the total number of sites, 
$f_{m(n)}$ is the Fermi-Dirac distribution function 
for energy $\mathcal{E}_{m(n)}$, 
$\ket{m}$ and $\ket{n}$ are single-particle eigenstates 
with energy $\mathcal{E}_m$ and 
$\mathcal{E}_n$ and $\eta$ is the scattering rate 
of conduction electrons from the localized spins. 
$v_x$ and $v_y$ are the velocity operators along $\hat{\mu} = \hat{x}, \hat{y}$ 
directions,
\begin{align}
v_\mu =\frac{i}{\hbar}\sum_{j,\mu,\sigma} [(t c_{j,\sigma}^\dagger c_{j+\hat{\mu},\sigma}- \mbox{H.c.})  + (t' c_{j,\sigma}^\dagger c_{j+\hat{\mu},\sigma}- \mbox{H.c.})]. 
\end{align}


\section{Results}
\label{sec:results}
	
\subsection{Magnetic properties}

\begin{figure}[b]
\centerline{
\includegraphics[width=7.0cm,height=3.5cm]{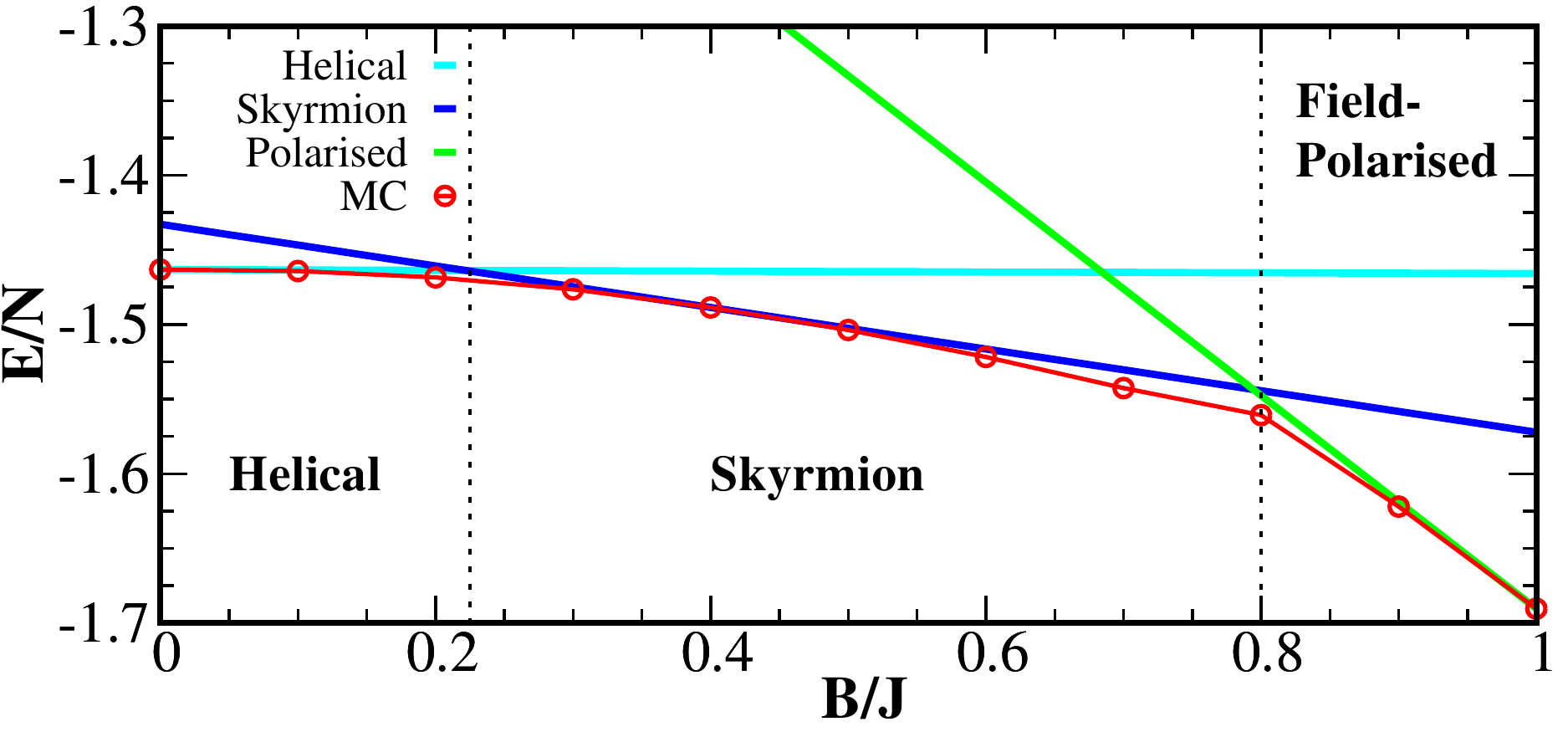}
}
\caption{\label{gs}
Color online: Energy per lattice site for different ordered phases 
with changing Zeeman field. This is compared with the energy of the configuration
obtained from the Monte Carlo simulation at the lowest temperature.
The ground state phases are determined by comparing the energies of 
the three static phases and the critical fields are identified as the points 
of energy level-crossing of the phases. The helical phase at $B/J=0.0$, 
the Skyrmion crystal phase at $B/J=0.4$ and a fully polarized phase 
are used as the reference states for carrying out the variational calculation. 
Comparison with the Monte Carlo data shows that the simulation results 
are consistent with the variational approach, providing an important confirmation that no possible phases are unaccounted for.
}
\end{figure}

\begin{figure*}[t]
\centering
{
\includegraphics[width=16.5cm,height=4.5cm]{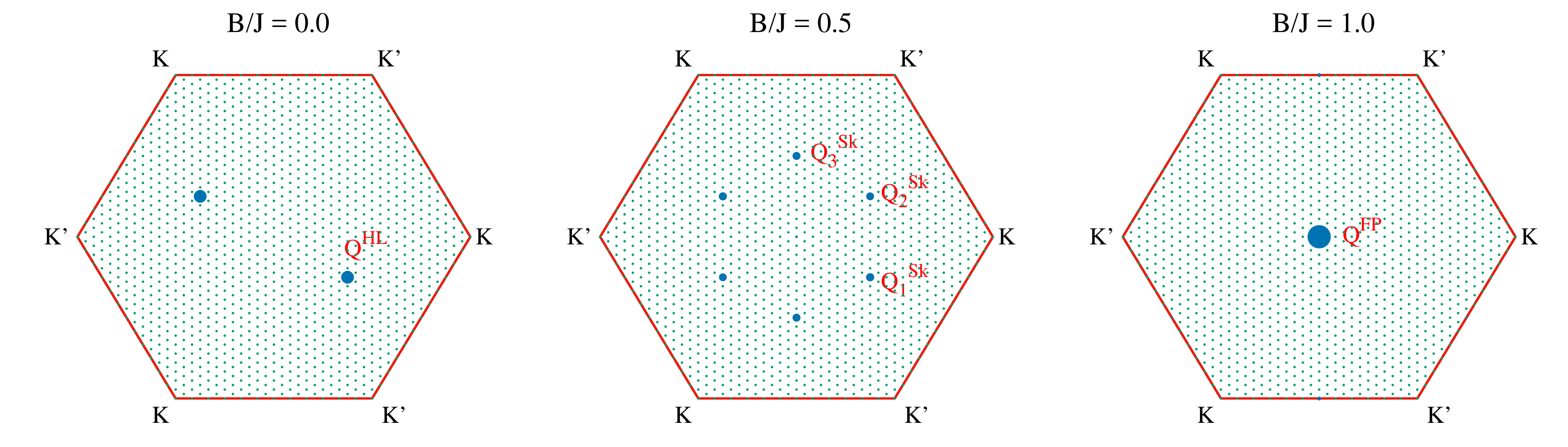}
}
\centering
{
\includegraphics[width=16.5cm,height=5.5cm]{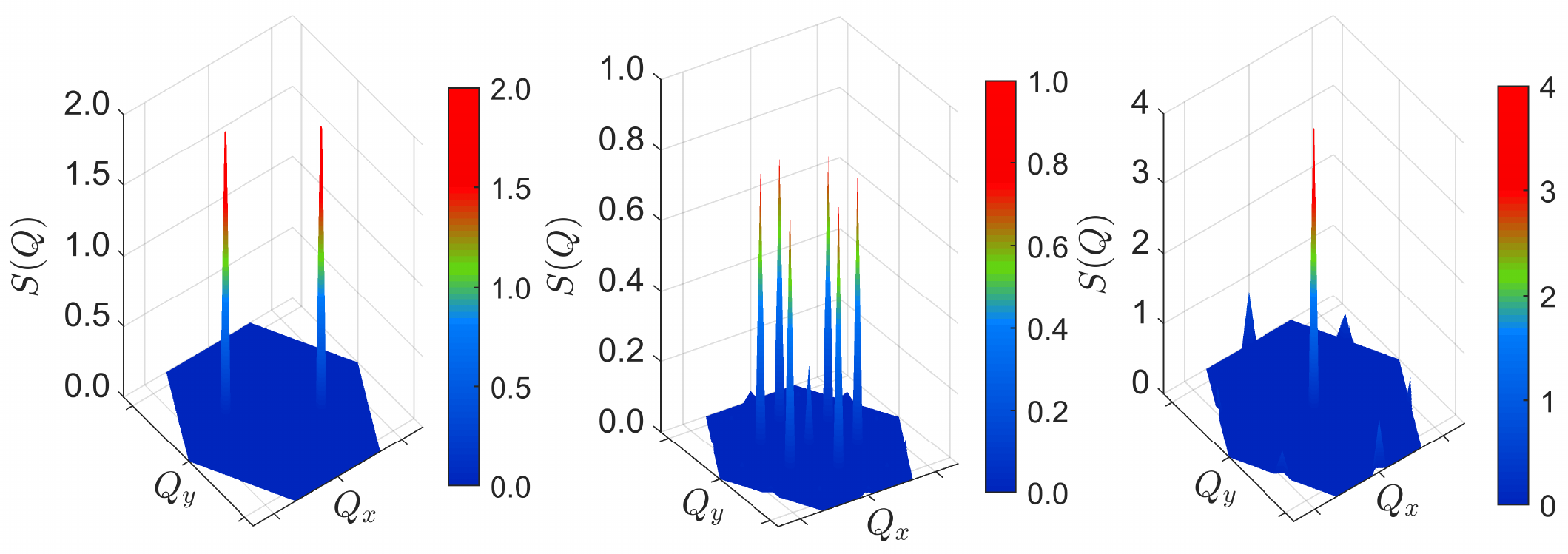}
}
\caption{\label{sf1}
Color online: Spin structure factor $S({\bf Q})$ in the ground state
at indicated Zeeman field values. 
(a)~$B = 0$ refers to the helical phase, where, $S({\bf Q})$
shows two prominent Bragg peaks (seen in bottom panel and as dominant weights in the color map) 
at ${\bf Q^{HL}} = \pm(\frac{2\pi}{4}, -\frac{2\pi}{4\sqrt{3}})$. 
(b)~The Skyrmion phase is seen at $B = 0.5J$, 
where $S({\bf Q})$ shows six prominent peaks 
in the $Q_{x}$ and $Q_{y}$ plane. These peaks are at,
${\bf Q_1^{Sk}} = \pm(\frac{2\pi}{4}, -\frac{2\pi}{4\sqrt{3}})$,
${\bf Q_2^{Sk}} = \pm(\frac{2\pi}{4}, \frac{2\pi}{4\sqrt{3}})$, and
${\bf Q_3^{Sk}} = \pm(0, \frac{2\pi}{2\sqrt{3}})$. 
In addition, we observe a weak peak at ${\bf Q} = 0$, 
the existence of which is attributed to the applied magnetic field along the longitudinal direction.
(c)~At $B = J$, $S({\bf Q})$ shows a prominent peak at ${\bf Q^{FP}} = (0,0)$
indicating the presence of a fully polarised phase.
}
\end{figure*}

{\bf Local spin configuration:} 
A snapshot of the ground state spin configuration from the simulations 
provides a visual insight into the nature of the magnetic phase.
We represent the localized spins as follows $-$ the planar components
$S_{ix}$ and $S_{iy}$ are represented by arrows in the $xy$ plane, 
whereas the out-of-plane component, $S_{iz}$, is represented by a color scale. Choosing representative values of $J_1$, $J_2$, $J_1'$ and $A$ that
are consistent with experimental observations in \ce{Gd3Ru4Al12} and varying the strength of the external field, $B$, we find three principal phases are stabilized -- the spiral phase at zero field with a periodicity of two unit cells, a Skyrmion crystal at intermediate fields and a fully polarized phase at high fields. 
Fig.\ref{config1} shows the local spin configurations for three distinct ground state phases.

We use a variational principle to determine the approximate phase boundaries between the field induced phases. Fig.\ref{gs} shows the energies of the helical, Skyrmion 
and spin polarized states as a function of the magnetic field.
The state with the lowest energy evolves from 
the helical phase at low fields, to a Skyrmion phase at intermediate field strengths, and eventually to the field polarized 
ferromagnetic phase at high fields. The approximate critical field strengths for the magnetic ground state phase transitions 
are inferred from the level-crossing of the energy for the different phases.
Further, we note that the energy of the phases obtained from Monte Carlo simulation (at the lowest temperature)
follows very closely the energy of the variationally obtained ground state phases;
thereby providing a benchmarking for the simulation process 
to reach the ground state.

We observe the following features.
$(i)$~At low magnetic fields ($0\leq B/J \leq 0.225$), the ground state is in the helical phase.
With increasing temperature, thermal fluctuations randomize the spins,
and above a critical temperature the state becomes a paramagnet. 
$(ii)$~At intermediate field strengths ($0.225 \leq B/J \lesssim 0.8$) 
a Skyrmion phase is obtained as the ground state. 
The size of the Skyrmions observed in our simulations is rather small,
which agrees with the experimental observations in \ce{Gd}-based Skyrmion materials~\cite{Gd_expt2}.
However, this is in contrast to the observed sizes of Skyrmions 
in non-centrosymmetric chiral magnetic systems~\cite{Neubauer_THE_MnSi_2009,yu_FeGe_2011,wilhelm_FeGe_2011}. 
Further, at low temperature, the individual Skyrmions are arranged in a periodic manner
which is known as the Skyrmion crystal phase.
With increasing temperature, the periodic arrangement of Skyrmions 
is gradually lost to thermal fluctuations.
$(iii)$~At high field strengths ($B/J \geq 0.8$), 
the Skyrmion phase gets suppressed completely, 
and a ferromagnetic phase polarized along the field direction is obtained. 
Further increasing the magnetic field does not change the symmetry of this state.
This polarized phase begins to randomize at large temperature ($T \approx J$) regime. 
A quantitative characterization of these different phases is obtained from the
study of multiple observables as detailed below.


{\bf Structure factor:} A detailed understanding of the multiple
magnetic phases is provided by the spin structure factor, $S(\bf{Q})$,
which quantifies long range magnetic order in terms of prominent peaks 
(or dominant weight) in the momentum space. 
We observe the following features (See Fig. \ref{sf1}),
$(i)$~In the helical (low field) phase, $S({\bf Q})$
shows two prominent Bragg peaks (seen as dominant weights in the color map) 
at ${\bf Q^{HL}} = \pm(\frac{2\pi}{4}, -\frac{2\pi}{4\sqrt{3}})$. 
The two ordering momenta are not independent, but related by symmetry. 
This non-collinear ordered phase is a helical phase specified by wavevector ${\bf Q^{HL}}$.
$(ii)$~In the Skyrmion crystal phase, $S({\bf Q})$ shows 7-peaks 
in the $Q_{x}$ and $Q_{y}$ plane.
The peak at ${\bf Q} = (0,0)$ is a trivial one due to the uniform magnetization along the longitudinal direction.
The remaining peaks come in three pairs at 
${\bf Q_1^{Sk}} = \pm(\frac{2\pi}{4}, -\frac{2\pi}{4\sqrt{3}})$,
${\bf Q_2^{Sk}} = \pm(\frac{2\pi}{4}, \frac{2\pi}{4\sqrt{3}})$, and
${\bf Q_3^{Sk}} = \pm(0, \frac{2\pi}{2\sqrt{3}})$
where the momenta in each pair are related by symmetry. 
The Skyrmion state can be understood as a linear superposition of three spiral phases 
(each represented by two symmetry related wave-vectors)
and is termed a 3-{\bf Q} state. The sharp peaks in these
phase denotes a near-perfect close packed ordering of the Skyrmions. 
$(iii)$~In the field-polarized ferromagnetic phase, 
$S({\bf Q})$ shows a prominent peak at ${\bf Q^{FP}} = (0,0)$.

\begin{figure}[t]
\centerline{
\includegraphics[width=6.0cm,height=4.5cm]{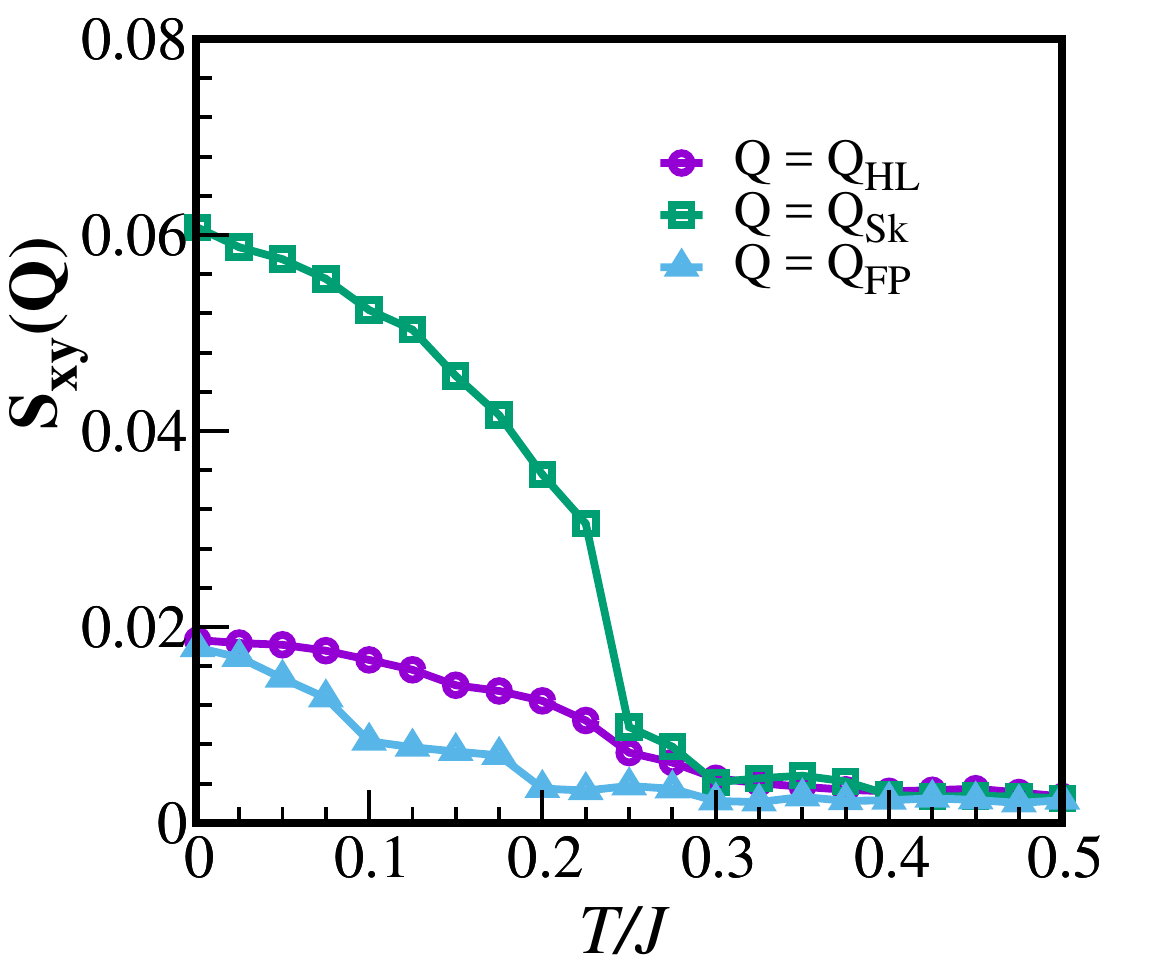}
}
\caption{\label{sf2}
Color online: Variation of the peak of the structure factor components 
$S_{xy}({\bf Q})$ with temperature for 
the helical, Skyrmion and the field-polarized phases respectively. 
The point of inflection separates the low-temperature phase 
from the high-temperature phase and represents the critical temperature ($T_c$)
for the respective phases. 
}
\end{figure}

A striking feature of the Skyrmions is that they are only two 
unit cells in size. This is in agreement with the experimental
observations of atomic scale Skyrmions in \ce{Gd2Ru4Al12}. 
Equally remarkable is the fact that it is achieved with no 
DMI and with  realistic microscopic 
interaction strengths. In contrast, in most chiral magnets
Skyrmions are realised in the presence of DMI whose strength
determines the size of the Skyrmions. While atomic scale 
Skyrmions are possible in principle via such a mechanism, the 
required strength (typically $D\gtrsim J$) is unrealistic 
in any naturally occuring chiral magnet. 
Large DMI can be engineered at magnetic interfaces in 
artificial heterostructures of ferromagnetic and heavy metal 
thin films, but synthesizing clean, defect-free samples with
a regular array of atomic scale Skyrmions is still not
possible with current technologies. 
On the other hand, DMI-free Skyrmions have been 
shown to arise from long range RKKY-type interactions 
in metallic magnets,~\cite{ozawa_zero-field_2017,skyrmion_RKKY_hexagonal}.

To check the stability of the magnetic phases against thermal fluctuations, we have studied the variation of the peak of the transverse structure factor
$S_{xy}({\bf Q})$ with increasing temperature (see Fig.\ref{sf2}).
Our results indicate that the non-coplanar spin textures are stable against small to moderate thermal fluctuations and the phases persist up to $T/J_1 \approx 0.25$. The structure factor weight reduces monotonically 
with increasing temperature across all phases.
This is expected as increased thermal fluctuations randomize the magnetic phases,
thus reducing the long-range correlation and eventually resulting in the loss of 
long-range order reflected in the complete suppression of the structure factor 
above a finite critical temperature. We estimate the critical temperature, $T_c$, as 
the point of inflection of structure factor weight variation with increasing temperature. 

\begin{figure}[b]
\centering
\includegraphics[width=6.5cm,height=5.0cm]{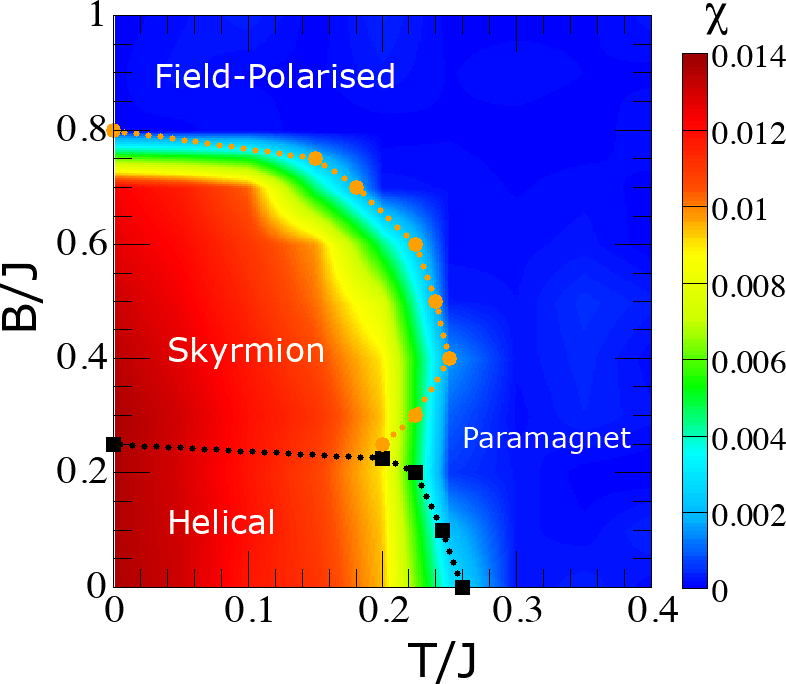}
\caption{\label{spin-chir}
Color online: Thermal phase diagram showing the behavior of spin chirality, $\chi$, 
obtained with our Monte-Carlo simulation for different Zeeman field values and temperatures. 
$\chi$ is non-zero in the helical phase and the Skyrmion phase. 
With increasing temperature, thermal fluctuations dominate, thereby
the ordered phases loose out to a paramagnetic phase with $\chi = 0$.
}
\end{figure}

{\bf Spin chirality:} One of the most interesting characteristics of 
complex non-coplanar spin textures is the non-zero chirality associated with them. 
To quantify the non-coplanarity of these spin textures 
we have calculated the scalar spin-chirality $\chi$ and studied 
its evolution with both temperature and applied magnetic field. 
Our results are summarized in Fig.~\ref{spin-chir}.
The chirality is non-zero in both the helical and the Skyrmion lattice phases reflecting the non-coplanarity of the spin textures in these phases, although the nature of non-coplanarity is different. 
The helical phase exhibiting finite spin chirality 
is likely due to the presence of
anisotropy.~\cite{Motome_review_multiQ_ordering_anisotropy}.

With increasing temperature,
thermal fluctuations destabilize the spin configurations;
the chirality decreases and eventually vanishes in the high temperature paramagnetic phase.
For $B/J \geq 0.8$, $\chi = 0$, the ground state has collinear ferromagnetic
ordering and the complete loss of non-coplanarity is reflected in a vanishing
chirality. 

To summarize, Hamiltonian (\ref{equ:ham-class}) exhibits a sequence of
non-collinear magnetic ground states in an external magnetic field. 
In the weak field regime, it gives rise to a helical phase; 
at intermediate field values, we observe a Skyrmion phase with atomic sized Skyrmions.  
Increasing temperature in this phase leads to a paramagnetic phase
above the critical temperature. 

\subsection{Electronic properties}

Coupling between itinerant electrons and local moments significantly
alters electronic transport properties 
in metallic magnets such as \ce{Gd4Ru2Al12}.
In the following, we 
explore the effects of such coupling on the transverse
conductivity in a magnetic field, including the emergence 
of a strong topological Hall effect.
For simplicity, we consider a single band of itinerant $s$-electrons interacting with 
the background spin configuration via a Kondo coupling 
between the electron spin and the local moments.
The dynamics of the electrons is fast compared to that of the localized classical spins. 
Consequently, at short time scales, the electrons effectively move in a static, 
but spatially varying magnetic field. 
Each local moment, ${\bf S}_i$ acts as a local magnetic field 
whose action on the spin magnetic moment of the itinerant electrons 
${\bf s}_i$ is described by a Kondo-like interaction 
$J_K{\bf s}_i\cdot{\bf S}_i$. 
In comparison, the Zeeman energy due the external magnetic field coupled to 
the spin of the electron is small and shall be neglected. Here, 
we discuss the effects of the different field induced spin textures on the band dispersion and topology, 
and the conductivity of the itinerant electrons. 

{\bf Band structure :}
The electron band structure on the regular kagome lattice consists of
three bands with 2-fold spin degeneracy, and a band width $w = 6t$. 
For a breathing kagome lattice, the size of the unit
cell is doubled. 
Coupling to the spin texture further increases the size of the unit cell in accordance with the magnetic unit cell which is
determined by the periodicity of the magnetic ordering. 
A non-zero $J_K$ lifts the spin degeneracy and the energy bands 
for electrons with spins parallel and anti-parallel to the local moments 
are shifted downwards and upwards, respectively. 
For sufficiently strong $J_K (> 6t)$, the two sets 
of bands are completely separated by a gap. For many metallic 
magnets, the dominant energy scale is this Kondo-like coupling
$J_K$ between the delocalized and localized degrees of freedom.
In the strong coupling limit, ($J_K \gg t$), 
the electron spins align with the local moments and the 
effective spatially varying magnetic field produced by 
the magnetic ordering couples directly to the charge of the 
electrons in a manner analogous to quantum Hall systems. 
The energy bands are renormalized by
the underlying spin texture and the Hamiltonian (\ref{equ:ham-elec}) reduces to an effective tight-binding model,

\begin{equation}
 {\mathcal{\hat{H}}_e}= -\sum_{\avg{i,j},\sigma}t_{ij}^{eff}(d_{i}^\dagger d_{j}+\mbox{H.c.}),
\end{equation}
where
\begin{equation}
t_{ij}^{eff}=t e^{ia_{ij}}\cos\frac{\theta_{ij}}{2},
\label{eq:teff} 
\end{equation}
is the effective hopping matrix for the spin-parallel electrons between sites $i$ and $j$ and the phase factor
\begin{equation}
a_{ij}=\arctan\frac{-\sin(\phi_i-\phi_j)}{\cos(\phi_i-\phi_j)+\cot\frac{\theta_i}{2}\cot\frac{\theta_j}{2}}
\end{equation}
that depends on the relative orientation of
the spins $\mathbf{S}_i$ and $\mathbf{S}_j$, 
and $\theta_{ij}$ is the angle between them. 
The spin anti-parallel electrons are described by 
a similar effective tight binding model with a different $t_{ij}^{eff}$ 
and the two sectors are completely decoupled.

\begin{figure}[b]
\centerline{
	\includegraphics[width=\linewidth]{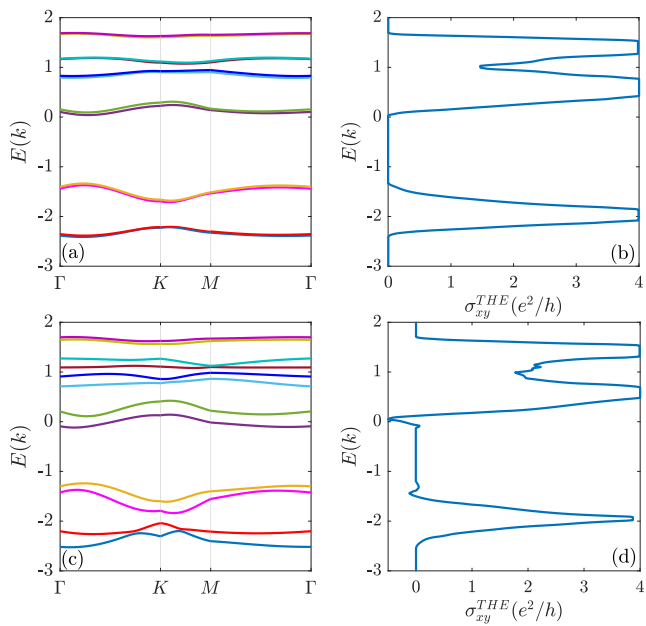}
}
\caption{\label{band-str}
Color online: (a),(c)~Band structure of conduction electrons moving on the backgrounds of
the Skyrmion phase and helical phase respectively on the breathing kagome lattice. 
Some bands carry non-zero Chern numbers. 
(b),(d)~Behavior of the topological Hall conductivity associated with 
the Skyrmion phase and the helical phase with varying chemical potential. 
}
\end{figure}

The topological character of an individual band is quantified  
by its associated Chern number, defined as 
$C_{n} = \frac{1}{2\pi} \int_{\mathrm{BZ}}\Omega_{n}^{(z)}({\bf k}) \mathrm{d}^{2}{\bf k}$
where $\Omega_{n}^{(z)}({\bf k})$ is the Berry curvature  given by
$\Omega_{n}^{(z)}({\bf k}) = 
\partial_{k_{x}} A_{n}^{(y)}({\bf k}) - \partial_{k_{y}}A_{n}^{(x)}({\bf k})$. 
In the above expression, 
${\bf A}_{n}({\bf k}) = 
-\mathrm{i} \langle u_{n}({\bf k})|\nabla_{{\bf k}}|u_{n}({\bf k}) \rangle$
is the Berry connection calculated from the eigenvectors $u_{n}({\bf k})$ 
with eigenvalues $E_{n}({\bf k})$ of the Hamiltonian~\ref{equ:ham-elec}.
We focus on the band structure in the spiral and Skyrmion lattice phases that exhibit non-coplanar magnetic orderings.

\begin{figure}[t]
\centerline{
	\includegraphics[width=6.5cm,height=5.5cm]{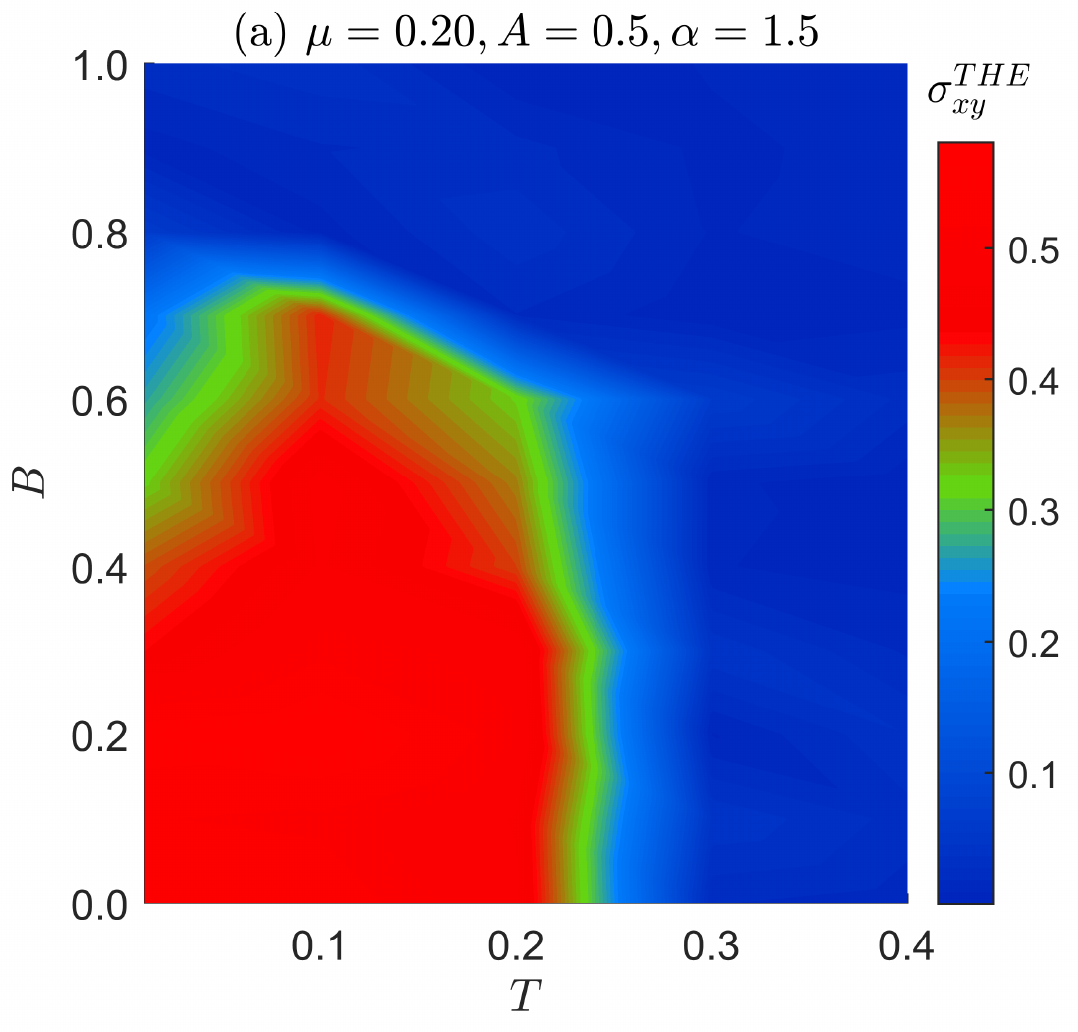}
}
\caption{\label{cond}
Color online: 
Behavior of $\sigma_{xy}$ at fixed chemical potential for varying temperature and magnetic field 
in the strong coupling limit ( $J_K \rightarrow \infty$).
}
\end{figure}

Interaction with local moments alters the 
transport properties of itinerant electrons in metallic magnets. The effect 
is most prominent in the transverse conductivity, especially for 
non-coplanar spin textures. In magnetic metals, the Hall resistivity consists of three contributions
\begin{equation}
    \rho_{xy} = \rho_{xy}^\text{NHE} + \rho_{xy}^\text{AHE} + \rho_{xy}^\text{THE}
\end{equation}
where NHE, AHE and THE refer to Normal, Anomalous and Topological
Hall effects, respectively. While the AHE arises due to spin-orbit coupling in metals with non-zero net magnetization, 
the THE is driven by the real space Berry phase 
acquired by an electron moving in a non-coplanar spin texture. 
The mechanism is best explained in terms of the effective 
Hamiltonian (\ref{equ:ham-elec}) in the
strong coupling limit ($J_K\gg t$). For the non-coplanar spin orderings, 
the local moments around a plaquette subtend a finite solid angle
at the center due to the  spatially varying spin texture. 
This results in a finite Berry phase when an electron hops around a plaquette and acts as a fictitious magnetic field 
with a flux $\Phi = (n_{sk}/\lambda^2)h/e$ through each plaquette. 
In the strong coupling limit, the phase of the effective hopping, 
$a_{ij}$, is associated with a vector potential 
acting on the itinerant electrons, analogous to quantum Hall systems. 
The dispersion of itinerant electrons is strongly affected by this Berry phase. 
The bands get narrower, a gap opens up between successive pairs of bands and 
each band acquires a finite Chern number (see fig. \ref{band-str}). 
Drawing on the analogy with quantum Hall systems, 
the bands can be described as dispersive Landau levels. The effective magnetic field drives a Hall effect, whose origin is purely geometric in nature.

In Fig.~\ref{band-str} we show the band structure for the itinerant electrons in the helical and the Skyrmion lattice phases.
We observe distinct features for the two different phases, with different Chern numbers of the bands.
At zero temperature only states below the Fermi energy, $E_f$ contribute to the transport.
When the Fermi energy lies within an energy gap, 
there is zero overlap between the current carrying states 
in the sample leading to the absence of backscattering processes. 
Thus the quantized value of $\sigma_{xy}$ signifies 
the absence of backscattering amongst the states and
a gap in the electronic energy spectrum. 
If $E_f$ is located within the band gap above any band, 
the topological component of the transverse conductivity, $\sigma_{xy}^{\text{THE}}$ is proportional to the the sum of the Chern numbers of all the filled bands.
Below the band gap, the Hall conductance decreases, while 
it increases above the band gap. 
This is due to the fact that the sign of the Berry curvature 
is opposite for the two adjacent bands.
This gives rise to a non-monotonic variation of $\sigma_{xy}^{\text{THE}}$, which is shown in
Fig.~\ref{band-str}(b) and (d).
The transverse conductivity exhibits several quantized plateaus (analogous to quantum Hall plateaus)
with increasing chemical potential. The jump in the conductivity is proportional to the Chern number of the bands.
The quantization of the Hall plateaus is most pronounced for small 
Skyrmion sizes where the lower bands are well separated. 
For larger Skyrmions, the density of the bands increases and the energy extent of the 
conductivity plateaus decreases proportionately as does the energy separation
between successive plateaus. This makes it difficult to resolve them in numerical simulations 
as effects of finite system size and fluctuations dominate. Importantly, the smaller Skyrmion size in \ce{GdRu4Al12} ensures that the mean free path of the electrons is comparable to, or larger than the Skyrmion size. This prevents the loss of coherence for the charge carriers due to scattering as they move through the Skyrmion texture and results in a giant THE as observed in experiments~\cite{Gd_expt4}.

As the magnetic field or temperature is varied, the transverse 
conductivity changes in accordance with the change in the underlying 
spin texture. Fig. \ref{cond} shows the variation of the topological 
Hall conductivity with temperature and applied magnetic field for a 
representative chemical potential. The helical and Skyrmion phases exhibit a 
strong THE which vanishes in the fully polarized  (with collinear 
magnetic order) and paramagnetic phase at strong magnetic fields
and high temperatures respectively. In the present work, we 
focus on isolating the THE contribution to the
transverse conductivity in the various field driven phases. 
We observe that $\sigma_{xy}$ is maximum in the low temperature regime, 
in the helical and Skyrmion phase. With increasing temperature, 
it reduces gradually and vanishes for $T > T_c$.

\section{Summary}\label{sec:summary}
We have studied the appearance of field induced Skyrmions 
in a breathing kagome lattice and its effects on electron transport.
Our results show that competing exchange interactions in association with anisotropies
and in the presence of external magnetic field gives rise to 
a Skyrmion phase in a centro-symmetric magnetic system.
The nature of different magnetic phases and 
the associated phase transitions are analyzed in detail 
with varying magnetic field and temperature. 
We have also studied in detail the topology of electronic band structure due to these
spin textures and reslting topological Hall effect seen in these systems, 
their variation with changing magnetic field, temperature,
carrier density and the tuning of the coupling between itinerant electron and localized spin. 
The effects of finite band dispersion on transverse conductivity are also analyzed.
Our results conclusively explain the recent observations of a Skyrmion phase and THE
in a breathing kagome magnet. 
			
{\it Acknowledgement.}---We acknowledge use of the computational resources at the High Performance Computing Centre (HPCC) at NTU (Singapore) and the National Supercomputing Centre (NSCC) ASPIRE1 cluster (Singapore). P.S. acknowledges support from the Ministry of Education (MOE), Singapore, in the form of AcRF Tier 2 grant  MOE2019-T2-2-119.


\bibliographystyle{apsrev4-1}
\bibliography{kagome}

\end{document}